\documentclass[fleqn,twoside]{article}
\usepackage{espcrc2}

\usepackage{graphicx,amssymb}
\usepackage[figuresright]{rotating}

\newcommand{\eq}{\begin{equation}}
\newcommand{\en}{\end{equation}}
\newcommand{\eqa}{\begin{eqnarray}}
\newcommand{\ena}{\end{eqnarray}}

\newcommand{\AmS}{{\protect\the\textfont2
  A\kern-.1667em\lower.5ex\hbox{M}\kern-.125emS}}

\title{ 
\vspace{-35mm}
\rightline{\small RNCP-Th02017,~~HU-EP-02/34,~~ITEP-LAT/2002-17}
\vspace{25mm}
Calorons and BPS monopoles with non-trivial holonomy
in the confinement phase of SU(2) gluodynamics%
\thanks{Presented at LATTICE 2002, Boston, by M. M.-P.}
}
\author{E.-M. Ilgenfritz\address[RCNP]{Research Center for Nuclear Physics,
        Osaka University, Osaka 567-0047, Japan}%
        \thanks{E.-M.I. acknowledges support by the Ministry
        of Education, Culture and Science of Japan (Monbu-Kagaku-sho).},
        B.V. Martemyanov\address[ITEP]{Institute for Theoretical and
        Experimental Physics, Moscow 117259, Russia},
        M. M\"uller-Preussker\address[HUB]{Humboldt-Universit\"at zu Berlin,
        Institut f\"ur Physik, 10115 Berlin, Germany},
        S. Shcheredin\addressmark[HUB]\thanks{Supported by DFG GK 271.}
        and
        A.I. Veselov\addressmark[ITEP]\thanks{Supported
        by grants RFBR 02-02-1730 and 01-02-17456, INTAS 00-00111 and
        CRDF award RP1-2364-MO-02.}}

\begin{document}

\begin{abstract}
With the help of the cooling method applied to SU(2) lattice gauge theory
at non-zero $T \le T_c$ we present numerical evidence for the existence
of superpositions of Kraan-van Baal caloron (or BPS monopole pair)
solutions with non-trivial holonomy, which might constitute an
important contribution to the semi-classical approximation of the
partition function.
\vspace{1pc}
\end{abstract}

\maketitle

A large number of applications to hadronic phenomenology has
proven the instanton liquid model to be a powerful
non-perturbative calculational framework of QCD. Whereas fundamental
non-perturbative features like spontaneous chiral symmetry breaking and
the large $\eta^{\prime}$-mass can be understood 
in terms of instantons, quark confinement seems to be related to other
excitations like Abelian monopoles or center vortices.
That instantons constitute important contributions to the Euclidean
QCD vacuum transition amplitude has found support from
lattice investigations. What possibilities exist to extend
the present instanton model? 
On one hand, correlations among instantons have to be taken 
into account. On the other, the question arises, whether 
other extrema of the Euclidean action also have to be taken into
consideration in the semi-classical ansatz. What r\^ole do play 
extended monopoles? 

For $~T \ne 0~$ the semi-classical approach \cite{GPY} so far was based on
superpositions of time-periodic instantons, the so-called calorons 
\cite{HS}. These HS calorons have trivial holonomy, i.e. 
the Polyakov loop at spatial infinity takes values in the center of 
the gauge group. Recently, Kraan and van Baal (KvB) have constructed a 
more general class of periodic and stable solutions of the Euclidean 
Yang-Mills field equations exhibiting non-trivial holonomy \cite{KvB}. 
The HS calorons turn out to be a limiting case of KvB solutions.
The main feature of the $SU(N_c)$ KvB solutions, having topological charge
$~Q_t=\pm 1~$, is that they can dissolve into $N_c$ separate static lumps of
non-integer topological charge, each of them constituting a BPS
monopole. Taken as a background field for the Dirac fermion operator, 
they lead to zero-modes concentrated only at one of these BPS monopoles
\cite{GPGAPvB}, depending on the boundary conditions used for the
fermion field. These features can be used to identify KvB
solutions in smoothed lattice fields.

In this contribution we are discussing some new results (see \cite{IMMPSV}) 
of continuing lattice investigations presented also at 
previous LATTICE conferences. 
For pure Yang-Mills theory at non-zero temperature we identify
low-action semi-classical configurations which are the background fields 
for equilibrium gauge fields in the path integral representation of the 
partition function. We start from Monte Carlo (MC) generated 
lattice gauge fields. By standard relaxation ('cooling') we minimize 
the action down to metastable plateaus shown to be approximate 
solutions of the lattice field equations. We restrict 
ourselves to the $SU(2)$ case and use the Wilson action. The lattice size 
throughout this paper is $16^3 \times 4$, i.e. the deconfinement transition 
occurs at $~\beta \equiv 4/g^2 = \beta_c \simeq 2.29~$. We shall concentrate 
on the confinement phase. In contrast to our previous investigations we 
employ standard periodic boundary conditions (p.b.c.) both for the MC 
simulation and for cooling. For the action as a function of cooling 
iterations we observe shoulders or plateaus close to multiple values 
of the one-instanton action value $S_{\mathrm{inst}}=2 \pi^2 \beta$.
We stop cooling on plateaus when the second derivative of the action
changes its sign from positive to negative values. At these stages of 
cooling the gauge fields are investigated with respect to the 
action density, the naively discretized topological density 
$q_t(x) = - \frac{1}{2^9 \pi^2} \sum_{\mu,\nu,\rho,\sigma=\pm 1}^{\pm 4}
           \epsilon_{\mu\nu\rho\sigma}
           \mathrm{tr}\left[ U_{x,\mu\nu} U_{x,\rho\sigma} \right],$
the Polyakov loop variable
$L(\vec{x}) = \frac{1}{2} \mathrm{tr} \prod_{t=1}^{N_t} U_{\vec{x},t,4}$
and the eigenvalues $\lambda$ and eigenmode densities 
$\psi \psi^{\dagger}(x)$ of the non-Hermitean Wilson-Dirac operator
$D[U]$, with
$\sum_{y} D[U]_{x,y} ~\psi(y) = \lambda ~\psi(x)$.
We have applied the Arnoldi method to both cases of time-antiperiodic and 
time-periodic boundary conditions of the fermion fields.
The statistics for our investigations were $O(200)$ per $\beta$-value. 

First we investigate the plateau configurations at 
$S \simeq S_{\mathrm{inst}}$ and below. There we have seen three types of
non-trivial objects. First of all, with the highest frequency of 
$60$ to $70 \%$, we have found approximate solutions which show all 
indications for (anti)selfdual KvB solutions consisting of one or
two separate lumps of topological charge. Both cases are characterized by
opposite-sign peaks of the Polyakov loop and non-trivial ('asymptotic')
holonomy values. The Wilson-Dirac operator exhibits always 
one distinct real ('zero'-) mode. For time-antiperiodic b.c.
the related eigenfunction is localized just at the peak of the 
Polyakov loop which corresponds to so-called Taubes winding 
(see \cite{KvB}). Repeating the eigenmode investigation with periodic
b.c. provides eigenfunctions localized at the complementary Polyakov 
loop peak in exact correspondence with the analytic results 
\cite{GPGAPvB}. We have denoted these configurations as calorons 
($CAL$), if they were consisting of one topological lump and were non-static 
in the time-direction, whereas we called them dyon-dyon pairs ($DD$) if a 
dissociation into two lumps of (non-integer) topological charges became 
visible and their fields were static in time. 
\begin{figure}[!htb]
\vspace*{-0.5cm}
(a)\hspace{0.1cm}
\includegraphics[height=0.15\textwidth,width=0.4\textwidth]{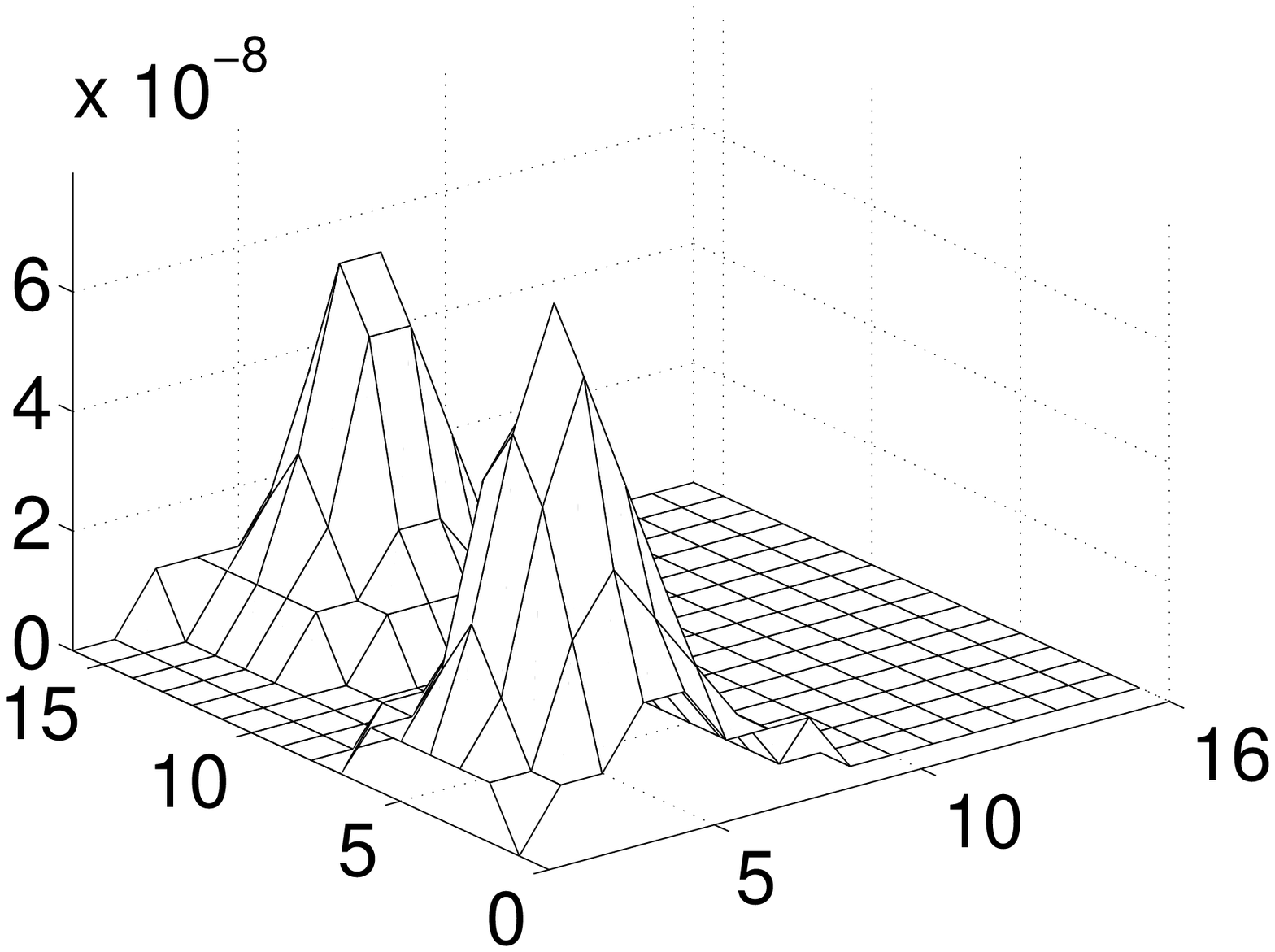}
(b)\hspace{-0.04cm}
\includegraphics[height=0.15\textwidth,width=0.42\textwidth]{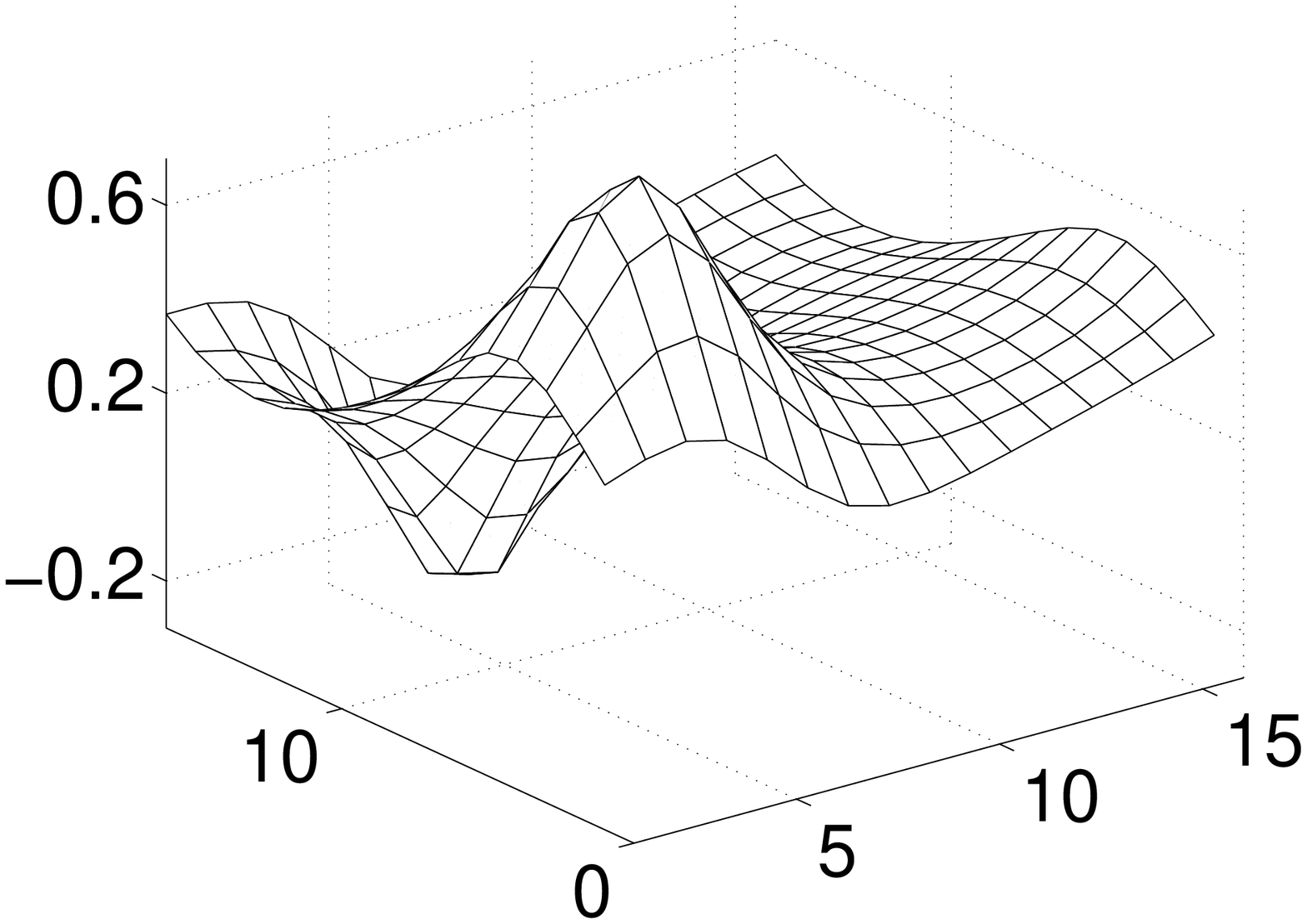}
(c)\hspace{0.5cm}
\includegraphics[height=0.15\textwidth,width=0.4\textwidth]{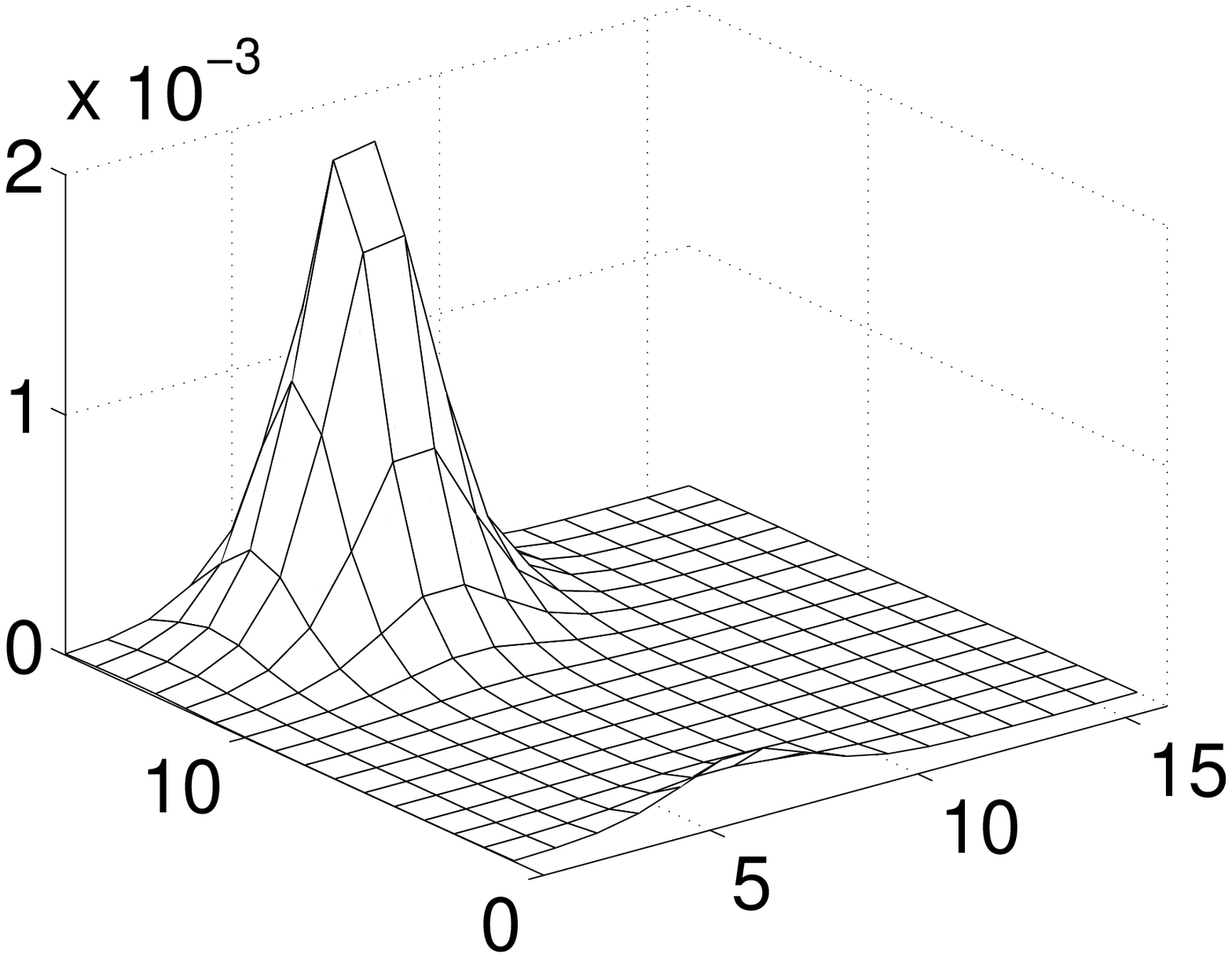}
\vspace*{-0.8cm}
\caption{
 $2D$ cuts for a $DD$ pair after cooling: 
 (a) topological charge density,
 (b) Polyakov loop,
 (c) 'zero'-mode density for time-antiperiodic b.c.
}
\vspace*{-0.5cm}
\label{fig:dd}
\end{figure}
A typical $DD$ configuration has been drawn in Fig. \ref{fig:dd}. 
Fits of $DD$ and $CAL$ configurations with the analytic KvB 
solutions were done and have shown to work well.
Therefore, we conclude that KvB solutions at lowest instanton action
plateaus are found dominating for $T \le T_c$.\footnote{The situation
is completely different for $T > T_c$. Self\-dual semi-classical 
fields become strongly suppressed under cooling if p.b.c. are applied.}
The second class of very stable configurations (around $10$ to $20 \%$)
were dyon-antidyon pairs ($D\overline{D}$) consisting of lumps of 
opposite half-integer topological charge and of equal-sign peaks 
of the Polyakov loop variable. The fermionic spectrum showed pairs 
of almost real eigenvalues complex conjugate to each other. 
These objects are not described by KvB solutions and look like peculiar
superpositions of BPS monopoles of opposite topological charge. Their
r\^ole is still unclear. Finally, in quite rare cases (less than $10 \%$)
we have seen pure magnetic objects like Dirac sheets or pure magnetic
monopoles. In contrast, such objects seem to dominate the semi-classical 
structure at $T > T_c$ \cite{LS}.    

Higher action plateaus $S \simeq (2 \cdots 8) S_{\mathrm{inst}}$
have been investigated for $T \le T_c$ with the same methods and observables.
One can easily identify superpositions of $CAL$'s, $D$'s and 
$\overline{D}$'s. The Polyakov loop variable as well as the fermionic 
modes with both kinds of boundary conditions allowed us to prove that these 
configurations always exhibit a pair structure (of Polyakov line peaks)
supporting the view of superpositions of KvB solutions with an admixture
of $D \overline{D}$-pairs. Of course, individual pairs are hardly 
identified.

\begin{figure}[!htb]
\vspace*{-0.5cm}\hspace*{0.5cm}
\includegraphics[height=0.15\textwidth,width=0.35\textwidth]{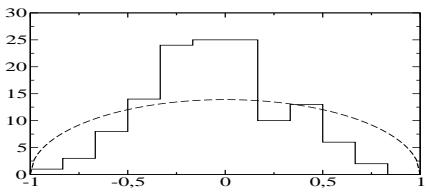}
\vspace*{-0.5cm}
\caption{
 Histogram $P(L_{\infty})$ 
 for $S \simeq 4 S_{\mathrm{inst}}$ plateaus 
 compared with the Haar measure (dashed line).
}
\vspace*{-0.5cm}
\label{fig:holonomy_dist}
\end{figure}

\begin{figure}[!htb]
\vspace*{-0.8cm}
(a) \hspace{0.05cm}
  \includegraphics[height=0.20\textwidth,width=0.4\textwidth]{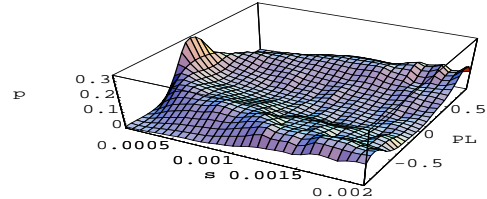}
(b) \hspace{0.05cm}
  \includegraphics[height=0.20\textwidth,width=0.4\textwidth]{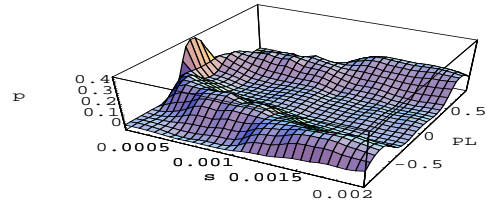}
\vspace*{-0.5cm}
\caption{
 Conditional distributions $P[L|\varsigma]$ 
 (a) at plateaus $S \simeq 4 S_{\mathrm{inst}}$,
 (b) for randomly distributed KvB solutions.
}
\vspace*{-0.5cm}
\label{fig:cond_dist}
\end{figure}

What is the typical holonomy ascribed to these configurations? 
We have determined the distribution $P(L_{\infty})$ of the Polyakov loop 
averaged over low action regions
$M = \left\{~\vec{x} ~~|~~ \frac{1}{N_t}\sum_t s(\vec{x},t) 
                                    \le 10^{-4} ~\right\}$
in the form
$L_{\infty}= \frac{1}{N_M} \sum_{\vec{x} \in M} L(\vec{x})$.
The result is seen in Fig. \ref{fig:holonomy_dist}. The histogram peaks 
around zero indicating the dynamical dominance of non-trivial holonomy.
The Polyakov line behaviour in different intervals of local action can be
used in order to see that also complicated superpositions 
of $D$'s and $\overline{D}$'s are similar to KvB-solutions.
For low-$L_{\infty}$ configurations we have found 
a local correlation between the Polyakov loop $L_(\vec{x})$ 
and the action density values 
$\varsigma(\vec{x})=\frac{1}{N_t}\sum_t s(\vec{x},t)$ in the form of
{\it conditional distributions} $P[L|\varsigma]$. 
In Fig. \ref{fig:cond_dist} we compare the result seen on 
four-instanton plateaus with that of lattice-discretized KvB solutions 
with randomly distributed parameters. 

We compared also with randomly distributed 'old-fashioned'
HS calorons of trivial holonomy. The latter provide a completely different 
picture. Therefore, the conditional distributions $P[L|\varsigma]$ 
are a hint in favor of KvB solutions, also for ensembles with several 
lumps of topological charge.

We would like to conclude that an improvement of the semi-classical caloron 
approach to the path integral at $0<T<T_c$ should take into account
superpositions of solutions with non-trivial holonomy.

\end{document}